\documentclass[preprint2]{aastex}

\newcommand{\av}{\ensuremath{A_{V}}}
\newcommand{\ji}{\textit{J}}
\newcommand{\hi}{\textit{H}}
\newcommand{\ksi}{\textit{K\ensuremath{_S}}}
\newcommand{\hii}{H\,$_{\rm{II}}$}
\newcommand{\kt}{\ensuremath{k_{\rm{B}}T}}
\newcommand{\lx}{\ensuremath{L_{\rm{X}}}}

\newcommand{\nh}{\ensuremath{N_{\rm H}}}

\slugcomment{}
\shorttitle{Hard X-rays from the Welch Ring}
\shortauthors{Tsujimoto et al.}

\begin{document}
\title{Hard X-rays from Ultra-Compact \hii\ Regions in W49A}
\author{M.~Tsujimoto\altaffilmark{1}, T.~Hosokawa\altaffilmark{2},
E. D. Feigelson\altaffilmark{3}, K. V. Getman\altaffilmark{3}, \& P.~S.~Broos\altaffilmark{3}}
\altaffiltext{1}{Department
of Physics, Rikkyo University, 3-34-1, Nishi-Ikebukuro, Toshima, Tokyo 171-8501, Japan.}
\altaffiltext{2}{National Astronomical Observatory of Japan, Osawa, Mitaka, Tokyo
181-8588, Japan.}  \altaffiltext{3}{Department of Astronomy \& Astrophysics, The
Pennsylvania State University, 525 Davey Laboratory, University Park, PA 16802.}

\begin{abstract}
 We report the \textit{Chandra} detection of hard X-ray emission from the Welch ring in
 W49A, an organized structure of ultra-compact (UC) \hii\ regions containing a dozen
 nascent early-type stars. Two UC \hii\ regions are associated with hard X-ray emission
 in a deep Advanced CCD Imaging Spectrometer image exposed for $\sim$96.7~ks. One of the
 two X-ray sources has no near-infrared counterpart and is extended by $\sim$5\arcsec,
 or $\sim$0.3~pc, at a distance of $\sim$11.4~kpc, which is spatially aligned with the cometary
 radio continuum emission associated with the UC \hii\ region. The X-ray spectrum of the
 emission, when fit with a thermal model, indicates a heavily absorbed plasma with
 extinction of $\sim$5$\times$10$^{23}$~cm$^{-2}$, temperature $\sim$7~keV, and X-ray
 luminosity in the 3.0--8.0~keV band of $\sim$3$\times$10$^{33}$~ergs~s$^{-1}$. Both the
 luminosity and the size of the emission resemble the extended hard emission found
 in UC \hii\ regions in Sagittarius B2, yet they are smaller by an order of magnitude
 than the emission found in massive star clusters such as NGC 3603 and the Arches
 cluster. Three possibilities are discussed for the cause of the hard extended emission
 in the Welch ring: an ensemble of unresolved point sources, shocked interacting winds
 of the young O stars, and a wind-blown bubble interacting with ambient cold matter.
\end{abstract}
\keywords{stars: early-type --- ISM: HII regions --- ISM: individual (W49A, the Welch
ring) --- X-rays: ISM --- radio continuum: ISM}

\section{INTRODUCTION}
One of the interesting differences in X-ray--emitting phenomena between low-mass and
high-mass star-forming regions is the existence of diffuse X-ray emission in high-mass
regions. Extended emission in the soft X-ray band has been reported in some young
massive clusters (e.g., M17 and the Rosette nebula; \citealt{townsley03}) and in nebulae
surrounding Wolf-Rayet stars (e.g., S-308 and NGC 6888; \citealt{chu03,wrigge05}). Such
emission is unlikely to arise from supernova remnants because these regions are expected
to be too young to host supernova explosions. The X-ray spectrum can be explained by
thermal plasma emission with a temperature of $\sim$10$^{6}$~K and total luminosity of
$\sim$10$^{33}$--10$^{34}$~ergs~s$^{-1}$. In the latter two regions, soft X-rays are
considered to be caused by a single Wolf-Rayet star, and hence the physical situation is
simpler than the former two regions, which harbor a number of such massive objects.

The temperature, luminosity, and the physical scale of $\sim$5--10~pc of the soft
diffuse emission can be explained by the wind-blown bubble model, in which the fast
winds from massive stars are thermalized by shocks colliding in the surrounding cold
matter. \citet{weaver77} presented an analytical self-similar solution for this model,
proposing that shock heating forms a bubble of hot gas that is observed as soft extended
X-ray emission. The gas is contaminated by metals as a consequence of mass loss from
massive stars, including main--sequence O-type stars and Wolf-Rayet stars. Such
contamination has been observed as nitrogen-enriched chemical abundances in the optical
spectra of some nebula (e.g., \citealt{esteban92}). In the soft diffuse emission from
S-308, \textit{XMM-Newton} measured elevated abundances consistent with those reported
for the cold surrounding nebula, supporting the wind-blown bubble model for the X-ray
emission \citep{chu03}. The bubble model can be similarly applied to explain the diffuse
X-ray emission around planetary nebulae (e.g., \citealt{dyson92,kastner01}).

Much hotter extended emission has been reported by \textit{Chandra} in other massive
star-forming regions, including NGC 3603 \citep{moffat02}, the Arches and the Quintuplet
clusters \citep{yusef-zadeh02,law04}, Sagittarius B2 \citep{takagi02}, RCW 38
\citep{wolk02}, and NGC 6334 \citep{ezoe06}. The hard diffuse emission in these samples
shows X-ray properties that differ from those of the soft diffuse X-ray emission. The typical
plasma temperature is higher by more than an order of $\gtrsim$10~$^{7}$~K. The typical
apparent size is much smaller of $\sim$0.1--1~pc. Some of these hard spectra may be
non-thermal (RCW 38 and NGC 6334; \citealt{wolk02,ezoe06}). Some show a strong
fluorescent line from neutral or low-ionized iron at $\sim$6.4~keV (Sagittarius B2, the
Arches cluster; \citealt{takagi02,yusef-zadeh02}), indicating that cold matter is
involved in the process.

The origin of the hard extended emission is less clear. \citet{canto02} presented an
analytic model to describe the winds from a massive cluster interacting with each other
and with the cold interstellar matter. Winds colliding with each other can give rise to
a higher temperature in shock plasma than the winds colliding in the cold matter at
rest. \citet{takagi02} speculated that the extended emission in Sagittarius B2 may be a
collection of unresolved point sources comprising of both low-mass and high-mass members
in the giant molecular cloud (GMC). Low-mass young stellar objects (YSOs) have been
known as strong hard X-ray emitters due to flares caused by magnetic reconnection
\citep{feigelson99}. It has recently been proposed that magnetic activity may be
responsible for the hard emission seen in some young O stars
\citep{schulz03,gagne05,stelzer05}. There is also a possibility that the hard extended
emission may be a special case of the wind-blown bubble employed to explain the soft
diffuse emission. None of these explanations have conclusive observational evidence.

Such a phenomenon is particularly interesting in ultra-compact (UC) \hii\ regions.  UC
\hii\ regions are ionized gas of a small size ($\lesssim$10$^{17}$~cm) and a high
density ($>$10$^{4}$~cm$^{-3}$), which are associated with massive forming stars still
embedded in their natal molecular clouds \citep{churchwell02}. Two equally important
energy sources governing the physics of UC \hii\ regions are strong ultraviolet (UV)
radiation and stellar winds from the forming massive stars \citep{capriotti01}. The
effect of the UV radiation, on one hand, is easier to assess, as the $\sim$10$^{4}$~K
plasma in equilibrium between the UV photo-ionization and the hydrogen recombination
can be accessed using interferometers via centimeter continuum free-free emission, which
penetrates through the extreme extinction (\av\ $=$ 10$^{3}$--10$^{4}$~mag) common among
these regions. The effect of the stellar winds, on the other hand, is poorly constrained
by observations and is often overlooked. If the hard extended emission is the outcome of
such winds, it will provide a unique observational tool to investigate the wind effects,
which will contribute to a complete understanding of the evolution and morphology of UC
\hii\ regions.

In this paper, we report the result of a deep hard X-ray study of a cluster of UC \hii\
regions in W49A, known as the Welch ring. At a distance of $\sim$11.4~kpc
\citep{gwinn92}, W49A is one of the most active massive star-forming sites in our Galaxy
\citep{alves03}. The total luminosity exceeds $\sim$10$^{7}$~$L_{\odot}$. The Lyman
continuum intensity of $\sim$10$^{51}$~s$^{-1}$ is equivalent to $\sim$100 O stars
\citep{conti02}. The Welch ring is the most curious object in the region, consisting
of a dozen UC \hii\ regions in an organized circular shape with a radius of $\sim$1~pc
\citep{welch87}. A dozen massive stars are being formed at the same time in a coherent
structure. Numerous maser sources have been discovered \citep{gwinn92,depree00},
indicating ongoing inflow and outflow motions. This structure is unique in our Galaxy
and affords opportunities to seek insights into massive star formation and the
interactions between massive stars and their environment. The present study is the first
in the hard X-ray band.

\section{OBSERVATIONS AND REDUCTION}
An X-ray imaging-spectroscopy observation was conducted using ACIS (Advanced CCD Imaging
Spectrometer; \citealt{garmire03}) on-board the \textit{Chandra X-ray Observatory}
\citep{weisskopf02}. The observation consists of two pointings, one for 33.5~ks
(ObsID--5893) on 2005 August 3--4 and the other for 63.1~ks (ObsID--6355) on 2005
August 4--5. Both were planned with identical aim points
(R. A.\,$=$\,19$^{\rm{h}}$12$^{\rm{m}}$41\fs2\ and
decl.\,$=$\,09$^{\rm{d}}$06$^{\rm{m}}$51$^{\rm{s}}$) and roll angles, resulting in a
combined data set that covers a $\sim$17\arcmin$\times$17\arcmin\ field for a total
integration time of $\sim$96.7 ks. ACIS is sensitive in the 0.5--8.0~keV energy band
with an energy resolution of $\sim$150~eV at 6~keV. The size of the point spread
function (PSF), represented by the 90\% encircled energy radius of a point source, is
$\sim$1 and $\sim$2\arcsec\ for 1.5 and 6~keV photons within 2\arcmin\ around the
optical axis. The faintest detected source has a flux of 
$\sim$10$^{-15.3}$~ergs~s$^{-1}$~cm$^{-2}$ (0.5--8.0~keV). The detector was operated
with the Timed Exposure mode with the nominal frame time. We used the Very Faint
telemetry mode to remove background events based on 5$\times$5 pixel charge
distributions.

\begin{figure}[hbtp]
 \figurenum{1}
 \plotone{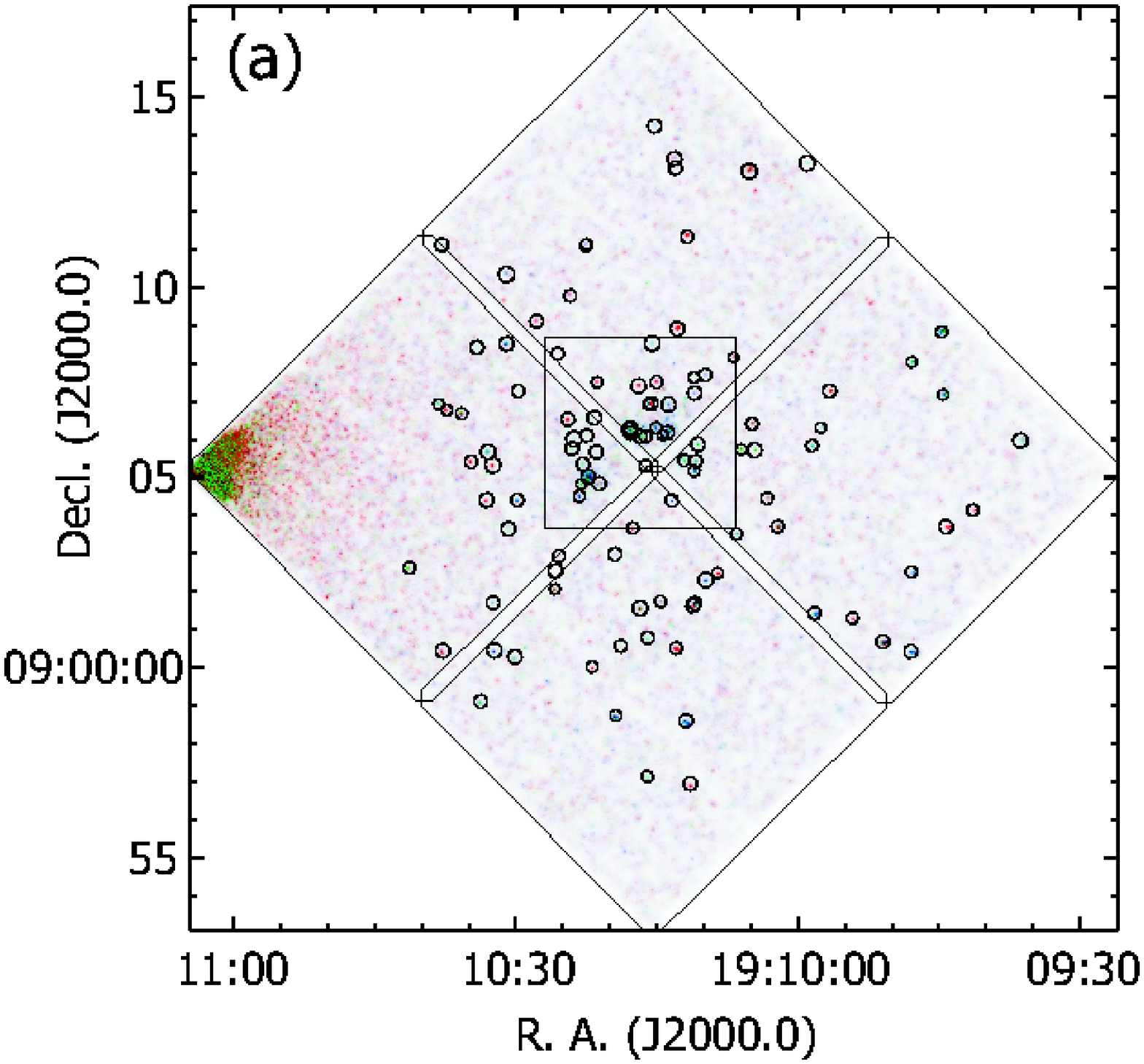}
 \plotone{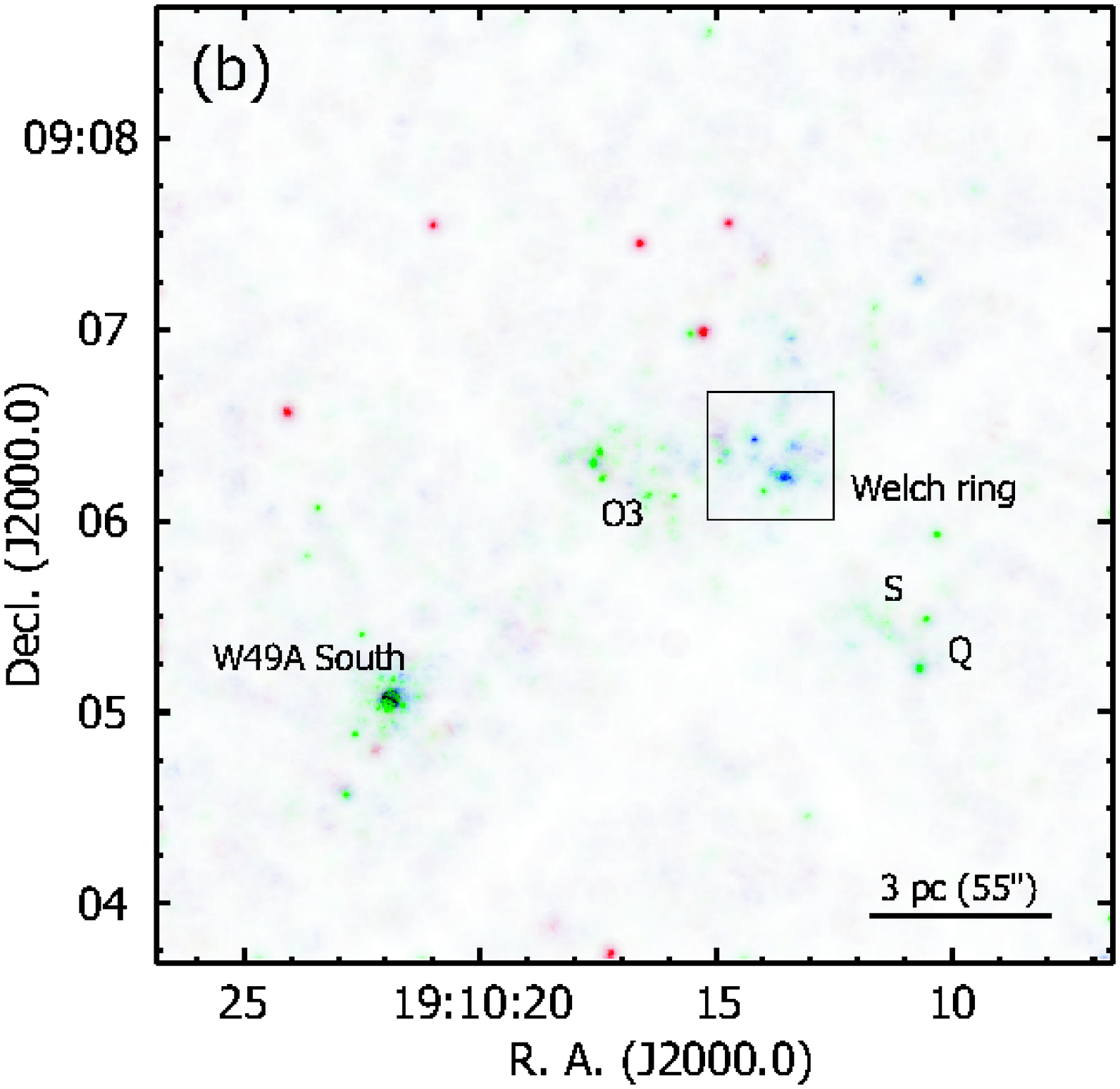}
 \caption{(\textit{a}) Pseudo-color smoothed image of the ACIS field. The views of the
 four CCD chips are shown with the oblique squares. Circles indicate the positions of
 detected X-ray sources. Red, green, and blue indicate the intensity in the 0.5--2.3
 (soft), 2.3--4.3 (medium), and 4.3--8.0~keV (hard) bands, respectively. The energy
 bands are chosen so that each band contains roughly equal numbers of
 photons. (\textit{b}) Close-up view of W49A (the central square in \textit{a}). The
 color coding is the same as (\textit{a}). The figure can be compared to Figure~1 in
 \citet{alves03} with the same dimension. An expanded view of the Welch ring region (the
 square in panel \textit{b}) is shown in Figure~\ref{fg:f2}.}\label{fg:f1}
\end{figure}

Figure~\ref{fg:f1}\textit{a} shows a pseudo-color smoothed image of the ACIS
field. The bright extended structure at the eastern edge of the field is the supernova
remnant W49B, which is unrelated to W49A. Figure~\ref{fg:f1}\textit{b} shows a closer
view of the major massive star clusters, including W49A South,
the Welch ring, O3, S, and Q. Near-infrared (NIR) sources associated with the clusters
W49A South, O3, S, and Q have colors redder than (\hi--\ksi) $=$ 2~mag due to a visual 
extinction of $\sim$32~mag \citep{alves03}, corresponding to a hydrogen-equivalent
column density of $\sim$5.7$\times$10$^{22}$~cm$^{-2}$ assuming a standard dust-to-gas
ratio \citep{predehl95}. Some X-ray photons are detected through this heavy extinction.

We used the CIAO package for the source detection and the
\anchor{http://www.astro.psu.edu/xray/docs/TARA/ ae_users_guide.html}{ACIS
Extract}\footnote{The ACIS Extract software and User's Guide have been available online
at \url{http://www.astro.psu.edu/xray/docs/TARA/ae\_users\_guide.html} since February
2003.} software package for the systematic analysis of X-ray photometry, variability,
and spectroscopy. The details of the procedure are described in \citet{getman05}. In
total, 107 X-ray sources were detected. The position of these sources are plotted in
Figure~\ref{fg:f1} (a). We concentrate on the hard X-ray emission associated with the
Welch ring in this paper. The X-ray properties of the other sources will be presented
separately. The astrometric uncertainty of the X-ray sources is $\sim$0\farcs35
estimated using 21 pairs of ACIS and Two Micron All Sky Survey (2MASS;
\citealt{cutri01}) counterparts.

\section{ANALYSIS}
\subsection{X-ray Sources in the Welch Ring}
\begin{figure}[hbtp]
 \figurenum{2}
 \plotone{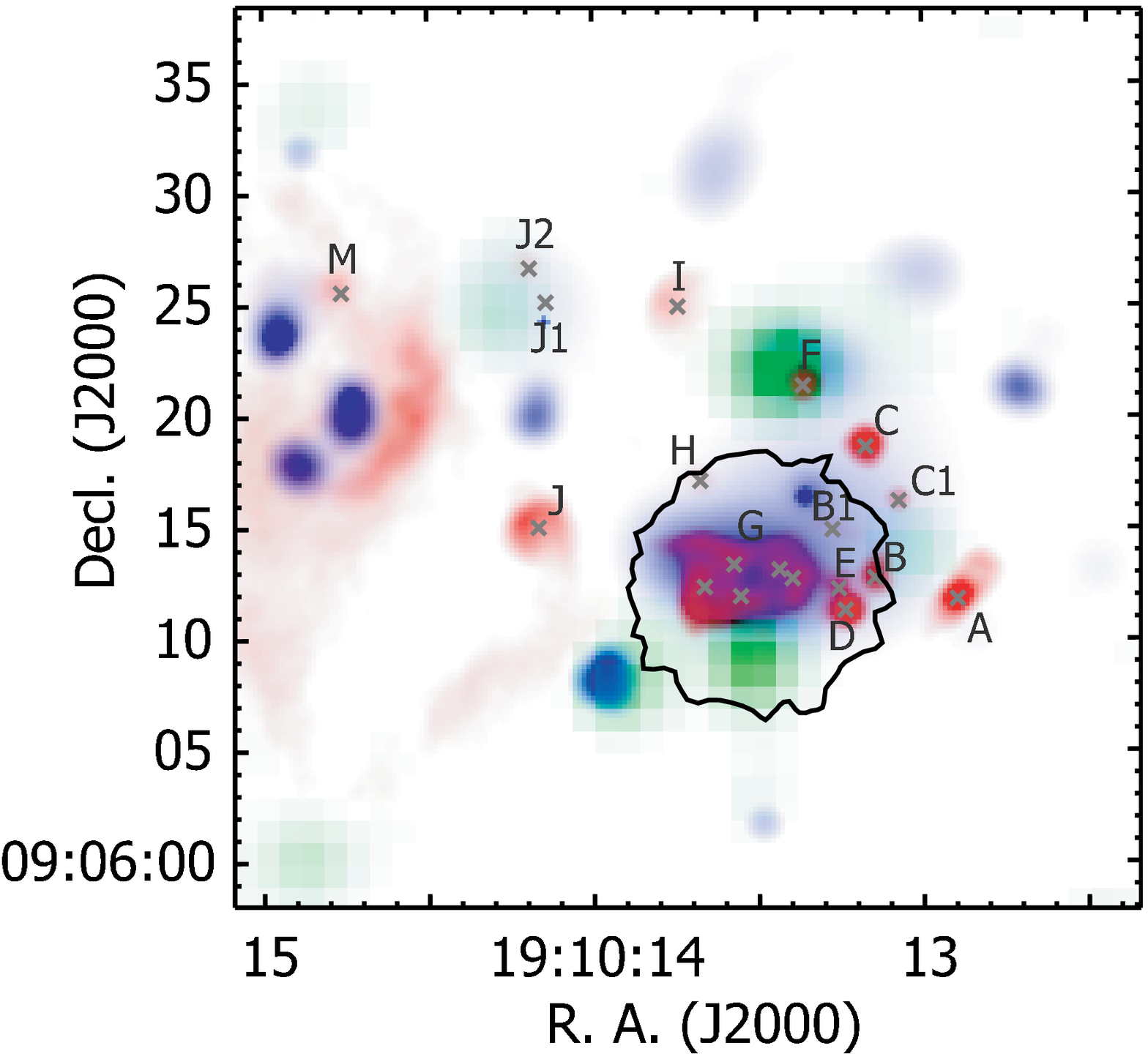}
 \caption{Multi-wavelength pseudo-color smoothed image of the Welch ring. Red, green,
 and blue indicate the intensity in the 3.6~cm \citep{depree00}, NIR \ksi\ (2MASS;
 \citealt{cutri01}), and the X-ray (0.5--8.0 keV) bands, respectively. Crosses indicate
 the positions of the 3.6~cm sources labeled with their names \citep{depree97}. The
 complex of objects named G1--G5 are labeled ``G''. The solid polygon is the extraction
 region of the X-ray events associated with the source G.}\label{fg:f2}
\end{figure}

Figure~\ref{fg:f2} shows a multi-wavelength pseudo-color smoothed image of the Welch
ring using X-ray, NIR, and radio continuum observations with blue, green, and red
colors, respectively. The ring can be best traced by the radio continuum image in red; a
dozen UC \hii\ regions labeled from A through M shape an ellipse of
$\sim$30\arcsec$\times$12\arcsec\ in major and minor axes and $\sim$60$\arcdeg$ in
position angle. The nomenclatures follow \citet{depree97}.

\citet{conti02} and \citet{alves03} reported NIR detections from two of the UC \hii\
regions (F and J2) identified by centimeter observations
\citep{dreher84,depree97}. Other NIR sources in Figure~\ref{fg:f2} are likely to be
foreground stars because of their blue colors in the \ji-, \hi-, and \ksi-band
photometry using the 2MASS data.

Among the UC \hii\ regions in the Welch ring, two are associated with X-ray emission (F
and G; Fig.~\ref{fg:f2}). From their positional coincidence, we consider that these two
X-ray sources are physically related to the UC \hii\ regions F and G. The X-ray source
at F is consistent with a point source, while that at G appears extended. This source is
the main subject of this paper; we designate it CXOU J191013.5$+$090612, hereafter
abbreviated to source G$_{\rm{X}}$.

\subsection{X-ray Image and Spectrum of G$_{\rm{X}}$}
\begin{figure}[hbtp]
 \figurenum{3}
 \plotone{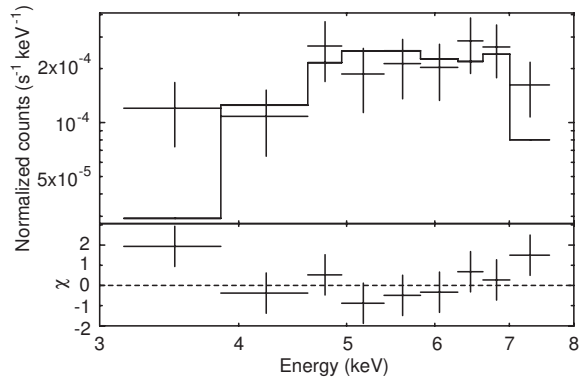}
 \caption{X-ray spectrum of the source G$_{\rm{X}}$. Grouped data with errors are shown
 in the upper panel over-plotted with the best-fit plasma model convolved with mirror
 and detector responses (\textit{solid line}). The lower panel shows the residuals of
 the fit.}\label{fg:f3}
\end{figure}

Figure~\ref{fg:f3} shows the spectrum of the source G$_{\rm{X}}$ extracted from the
aperture shown by the polygon in Figure~\ref{fg:f2}). For technical reasons, it was
convenient to make the aperture a contour of the PSF that would enclose 99\% of a
point-like source, which is large enough to cover the extended emission of
G$_{\rm{X}}$. The extracted events consist of 101 counts with an estimated local
background of $\sim$18 counts. No flux variability was confirmed with the null
hypothesis (constant flux) probability of 0.37. The spectrum was binned to 10
counts~bin$^{-1}$ and was fit over the energy range of 3.0--8.0 keV. The spectrum is
strongly cutoff at the softer end but is extremely hard such that the count rate hardly
falls despite the rapidly falling sensitivity of ACIS above $\sim$5 keV. This is very
unusual, indicating an extremely hard intrinsic spectrum that cannot be attributed to
heavy absorption. The median energy of source is 5.7 keV, which is matched by only 2
of the 1616 sources in the deep X-ray census of the Orion Nebula region
\citep{getman05}. Both thermal and non-thermal models were tried for fitting since the
data are not sufficient to detect or reject iron emission lines, which would otherwise
constrain the spectral models.

First, a thin-thermal--plasma APEC model \citep{smith01} with interstellar extinction was
applied. The metallicity was fixed to 0.3 solar following the convention in
\citet{getman05}. The minimum $\chi^{2}$ value was obtained for a hydrogen-equivalent
column density of \nh $=$ 5.1$^{+2.5}_{-1.8} \times 10^{23}$~cm$^{-2}$ and a plasma
temperature of \kt $\sim$ 7~keV. The fit is statistically acceptable, corresponding to a
null hypothesis probability for the $\chi^{2}$ value of 0.23. The uncertainties quoted
for the \nh\ value indicate the 90\% confidence interval. No meaningful confidence
interval was available for \kt\ because of the relatively poor spectrum. Values of
$\chi^{2}$ were calculated at 10$^{4}$ different sets of parameters in the range of
\nh$=$0--100$\times$10$^{22}$~cm$^{-2}$ and \kt$=$0.5--20~keV, but no other local
minimum was found. The flux in the 3.0--8.0~keV is $\sim$4 $\times
10^{-14}$~ergs~s$^{-1}$~cm$^{-2}$, and the absorption-corrected luminosity in the band is
$\sim$3$\times$10$^{33}$~ergs~s$^{-1}$.

Second, a power-law model of $E^{-\Gamma}$ with interstellar extinction was applied. The
minimum $\chi^{2}$ value was obtained for \nh $\sim$ 2.1$\times 10^{22}$~cm$^{-2}$ and
$\Gamma \sim$ --2.7. No other local minimum was found in the (\nh--$\Gamma$) plane. The
fit is statistically acceptable, but the positive power--law slope is
unphysical. Therefore, we hereafter refer to the results of the thermal fit.

\begin{figure}[hbtp]
 \figurenum{4}
 \plotone{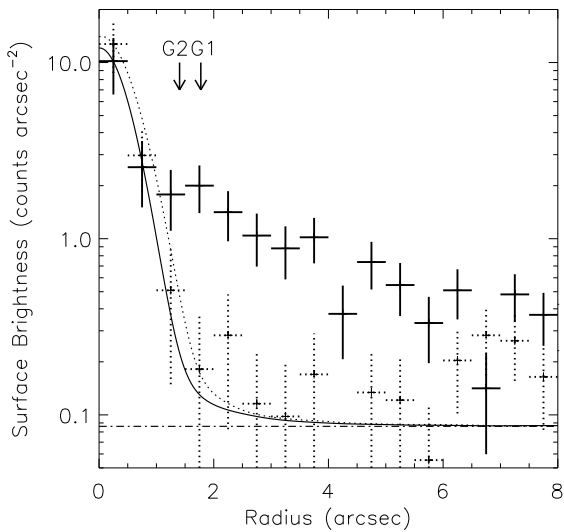}
 \caption{Observed radial profiles of the source G$_{\rm{X}}$ (\textit{solid pluses})
 and the comparison source CXOU J191010.4$+$090528 (\textit{dotted pluses}). The radial
 profiles of the local model PSFs (\textit{solid and dotted curves}) are shown, elevated
 on the observed background (\textit{dashed-and-dotted line}). The displacements from
 the center of G$_{\rm{X}}$ to the closest radio compact sources (G1 and G2;
 \citealt{depree97}) are shown by arrows.}\label{fg:f4}
\end{figure}

In order to examine the extent of the source G$_{\rm{X}}$ in comparison with the PSF, we
plot the radial profile of this source (Fig.~\ref{fg:f4}). A comparison source (CXOU
J191010.4$+$090528) at a similar off-axis angle with a similar median X-ray
energy is also plotted. G$_{\rm{X}}$ and the comparison source are located at
off-axis angles of 1\farcm6 and 1\farcm8 and have median energies of 5.7 and 4.1 keV,
respectively. We find that G$_{\rm{X}}$ is extended with a radius of $\sim$5\arcsec\
($\sim$0.3~pc) in contrast to the point-like comparison source, which has a profile
consistent with the PSF at its position.

\subsection{Source G in Other Wavelengths}
\begin{figure}[hbtp]
 \figurenum{5}
 \plotone{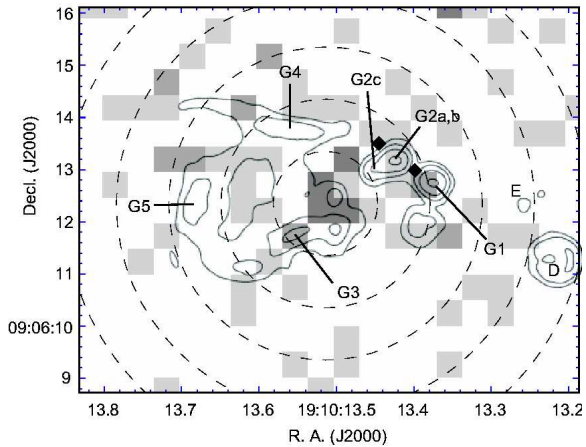}
 \caption{X-ray intensity map of the source G$_{\rm{X}}$ in grey scale (from 0 to 3
 counts~pixel$^{-1}$). The contours are 1.3~cm intensity (Fig.~1c in \citealt{depree00})
 with the names of radio compact sources (G1, G2a, and G2b) and knots in a lobe-like
 extended structure (G3--5). The concentric circles indicate the distance from the peak
 of the X-ray emission in 1\arcsec\ steps. The diamonds show the position of two
 additional peaks in the deconvolved X-ray image.}\label{fg:f5}
\end{figure}

In the region around G$_{\rm{X}}$, \citet{depree97} detected five sources (G1--5)
using the Very Large Array (VLA) in its shorter baseline configurations (C and D) at
3.6, 1.3, and 0.7~cm. \citet{depree00} subsequently detected G1 and resolved G2 into
three compact sources (G2a--c) using VLA in the longer baseline configurations (A and B)
at 1.3 and 0.7~cm. BIMA observations also detected the four compact sources (G1 and
G2a--c) at 3.3~mm \citep{depree00} and 1.4~mm \citep{wilner01} with a similar
synthesized beam width (0\farcs1--0\farcs3). These compact sources are presumably newly
formed massive stars. In contrast, the lack of detection of G3--5 by the higher
resolution interferometer observations indicates that these sources are parts of a
cometary structure, as is discussed in \citet{depree00}, extending $\sim$4\arcsec\
eastward of G1 and G2, shown as contours in in Fig.~\ref{fg:f5}. (See also see Fig. 1 in
\citealt{dreher84} and Fig. 1 in \citealt{depree00}). The nature of the cometary
emission is uncertain.

In the infrared regime, the extreme extinction to this source prevents access by NIR
\citep{conti02,alves03} and mid-infrared (MIR; \citealt{smith00}) observations shortward
of 12.3 \micron. In the 20.6 \micron\ image, a source appears at G, showing an extended
structure with a scale of $\sim$5\arcsec\ attributable to thermal emission from heated
dust grains \citep{smith00}.

The peak of the X-ray emission does not coincide with the positions of any compact
sources in the radio continuum (G1 and G2; Figs.~\ref{fg:f4} and \ref{fg:f5}), which are
displaced from the X-ray peak by $\gtrsim$\,1\farcs5 (Fig.~\ref{fg:f4}). We constructed
an X-ray image deconvolved with the PSF using a maximum likelihood reconstruction
technique \citep{lucy74}. Deconvolved images often introduce artificial peaks, and the
only practical way to examine the validity of these new peaks is to see whether they
match the sources detected in higher resolution images in other wavelengths. In the
deconvolved X-ray image around G$_{\rm{X}}$, we found two peaks at the positions shown
with the diamonds in Figure~\ref{fg:f5}. The locations of these peaks near the positions
of G1 and G2 suggest X-ray emission from these sources, especially if an astrometric
misalignment of $\sim$~0\farcs4 is allowed. A deeper observation will be required to
confirm emission from G1 and G2. However, there is no question that most of the hard
X-ray emission from G$_{\rm{X}}$ arises from a region displaced 1--2\arcsec\ eastward of
G1 and G2. No more than 10\% of the counts are from G1 and G2.

\section{DISCUSSION}
\subsection{Comparison with Other Hard Extended Emission}
We can summarize the X-ray properties associated with the UC \hii\ region G in the Welch
ring as follows. It has a luminosity of $\sim$3$\times$10$^{33}$~ergs~s$^{-1}$ and its
spectrum can be fitted with thermal plasma of a temperature of $\sim$7~keV and an
interstellar extinction of $\sim$5$\times$10$^{23}$~cm$^{-2}$. The profile is extended
and a size of $\sim$0.3~pc. The center of the diffuse emission is offset from the
closest compact radio sources.

These findings can be compared with the properties of the hard diffuse emission
previously claimed in other regions, listed in Table~\ref{tb:t1} in increasing order
of distance. The parameters of the diffuse emission in the Quintuplet cluster were not
constrained very well because the emission is overwhelmed by the intense diffuse
emission in the Galactic center \citep{law04}. The numbers in Table~\ref{tb:t1} cannot
be compared naively because of the different reduction techniques used by different
authors. However, the X-ray properties of G$_{\rm{X}}$ resemble those found for UC \hii\
regions in Sagittarius B2, in terms of the X-ray luminosity, the temperature, the amount
of extinction, and the size. Both the X-ray luminosity and the size of these sources are
smaller by an order of magnitude than those found in star clusters like NGC 3603, Arches, and
Quintuplet. It thus appears that there are two categories for the hard diffuse emission from
the view point of the X-ray luminosity and the size.

\begin{deluxetable}{lccccccl}
 \tabletypesize{\scriptsize}
 \tablecaption{Hard Diffuse X-ray Emission from High-Mass Star-Forming Regions.\label{tb:t1}}
 \tablecolumns{10}
 \tablewidth{0pt}
 \tablehead{
 \colhead{Region} &
 \colhead{\lx\tablenotemark{a}} &
 \colhead{\kt\tablenotemark{b}} &
 \colhead{\nh\tablenotemark{c}} &
 \colhead{Size\tablenotemark{d}} &
 \colhead{Distance} &
 \colhead{Comment} &
 \colhead{Reference} \\
 \colhead{} &
 \colhead{(ergs~s$^{-1}$)} &
 \colhead{(keV)} &
 \colhead{(cm$^{-2}$)} &
 \colhead{(pc)} &
 \colhead{(kpc)} &
 \colhead{} &
 \colhead{}
 }
 \startdata
RCW 38 \dotfill      & 1$\times$10$^{32}$ & 2.2 & 1$\times$10$^{22}$ & 1.5 & 1.7 & cluster & \citet{wolk02} \\
NGC 6334 \dotfill    & 0.1--5$\times$10$^{32}$ & $>$1 & 0.5--10$\times$10$^{22}$ & 1 & 1.7 & GMC & \citet{ezoe06} \\
NGC 3603 \dotfill    & 2$\times$10$^{34}$ & 3.1 & 7$\times$10$^{21}$ & 4 & 7.0 & cluster & \citet{moffat02} \\
Arches \dotfill      & 2$\times$10$^{34}$ & 5.7 & 1$\times$10$^{23}$ & 3 & 8.5 & cluster & \citet{yusef-zadeh02} \\
Quintuplet \dotfill  & 1$\times$10$^{34}$ & \nodata & \nodata & \nodata & 8.5 & cluster & \citet{law04} \\
Sgr B2 source 10 \dotfill & 9$\times$10$^{32}$ & 10 & 4$\times$10$^{23}$ & 0.2 & 8.5 & UC \hii & \citet{takagi02} \\
Sgr B2 source 13 \dotfill & 1$\times$10$^{33}$ & 5 & 4$\times$10$^{23}$ & 0.2 & 8.5 & UC \hii & \citet{takagi02} \\
W49A G$_{\rm{X}}$ \dotfill & 3$\times$10$^{33}$ & 7 & 5$\times$10$^{23}$ & 0.3 & 11.4 & UC \hii & this work \\
 \enddata
 \tablenotetext{a}{Absorption-corrected X-ray luminosity.}
 \tablenotetext{b}{Plasma temperature.}
 \tablenotetext{c}{Interstellar extinction.}
 \tablenotetext{d}{Physical size of the diffuse emission.}
\end{deluxetable}

Based on the results in the Welch ring and the comparison with the other regions, we
explore the three possible causes outlined in \S 1: a collection of unresolved point
sources, interacting O star winds, and a wind-blown bubble.

\subsection{Population of  Point Sources}
The first possibility for the cause of the diffuse emission at G is that it is 
a collection of unresolved point sources. As the region contains forming early-type
stars, it should also contain numerous low-mass YSOs born from the same
molecular cloud. Because both low-mass YSOs and massive young stars show hard X-ray
emission, the collection of these sources can mimic hard diffuse emission when they are
unresolved. The total X-ray emission from these sources can be estimated in the following
manner.

First, the number of massive stars in this region is inferred using two indirect
methods; any direct census of massive stars is difficult due to the extreme extinction
that prevents observations of photospheric emission from these stars in the optical and
NIR bands. One method is to use the luminosity of dust emission warmed by massive stars
in the MIR band. \citet{smith00} found that the integrated luminosity between 12 and
20 $\mu$m is $\sim$2$\times$10$^{5}~L_{\odot}$, which corresponds to a single zero-age
main-sequence (ZAMS) star of spectral type O6. Another method is to use the Lyman
continuum photons emitted per unit time derived from the radio continuum
flux. \citet{depree00} estimate that two O5.5 ZAMS stars are present. These two
independent estimates both indicate there are 1--2 O5--6 ZAMS sources in region G.

We scale the estimate of high-mass population in G$_{\rm{X}}$ to the X-ray luminosity
function of the Orion Nebula Cluster (ONC), the star-forming region best studied by
\textit{Chandra} \citep{getman05}. In the ONC, one O6 V star ($\theta^{1}$ Ori\,C)
accounts for one third of the total hard emission, while the whole of the lower mass
(0.3--3.0~$M_{\odot}$) YSO population accounts for one half \citep{feigelson05}. In
G$_{\rm{X}}$, where 1--2 O5--6 sources are present, we estimate that the total hard
X-ray luminosity of point sources is $\sim$1--2$\times$10$^{33}$~ergs~s$^{-1}$, which is
on the same order as the observed luminosity.

In NGC 3603 and NGC 6334, \citet{moffat02} and \citet{ezoe06} respectively estimated
that $\sim$25\% and $\sim$20\% of the diffuse flux can be explained by unresolved faint
point sources based on the X-ray luminosity function constructed in each region. In the
Welch ring, on the other hand, we cannot construct the luminosity function specific to
this region. Considering this ambiguity, it is possible that all of the X-ray flux from
G$_{\rm{X}}$ can be attributed to the integrated emission of unresolved point sources.

\subsection{Interacting Winds}
In massive star clusters, winds from each massive star collide to produce shock-heated
plasma \citep{ozernoy97}. This is a favored model to explain the extended hard X-ray
emission based on the fact that it is easier to produce hard X-ray emission because it
can produce higher temperature plasma than the bubble model (\S 4.4). In the Welch ring,
the interacting winds could naturally explain the fact that similar extended X-ray
emission is not found in other UC \hii\ regions at the same level; G is the only region
that was resolved into several bright compact radio continuum sources \citep{depree00},
indicating the existence of multiple early-type stars. Numerical models have been
developed to explain the diffuse emission associated with the Arches cluster
\citep{raga01,canto02,rockefeller05}. A similar model could be developed for the Welch
ring emission for quantitative comparison.

A related possibility is the production of hard X-rays from small-scale wind shocks in
Wolf-Rayet binary systems. Here, the X-ray luminosity reaches $\sim 10^{33}$ erg
s$^{-1}$ and temperatures as high as $\sim$10$^{8}$~K (e.g.,
\citealt{pollock87,usov92}). Most of the observed flux should arise from the position
where the two winds collide head-on to produce a peak temperature plasma; plasma in the
periphery is subject to cooling as discussed in the next subsection. The observed
luminosity and temperature of G$_{\rm{X}}$ are similar to those observed in Wolf-Rayet
binary systems. Therefore, the X-ray emission in G can also be explained by the
colliding winds from multiple forming massive stars.

\subsection{Wind-blown Bubble}
A wind-blown bubble model has been employed to explain the soft diffuse emission in
other regions \citep{townsley03,chu03,wrigge05}. The application of this model to
G$_{\rm{X}}$ is inspired by the fact that G, unlike other UC \hii\ regions in the
Welch ring, is also associated with lobe-like extended radio emission with the same
scale around the diffuse X-ray emission. The bubble model has also been used to explain the
coherent space velocities of water maser sources from the very source of G
\citep{maclow92}, where it was proposed that these sources are from the expanding shells
swept by stellar winds. The involvement of cold matter in the hard X-ray emission is
strongly inferred by the detection of the fluorescent iron line in Sagittarius B2 and
the Arches cluster \citep{takagi02,yusef-zadeh02}. Because this possibility for the
X-ray emission has not been discussed in previous works, we examine this idea here in
more detail.

Wind-blown bubbles are formed as a consequence of strong stellar winds from massive
stars and have been modeled in various circumstances. A classic study of
\citet{weaver77} discusses a case in which the strong wind collides with the uniform
cold interstellar medium of a $\sim$1~cm$^{-3}$ density. They argued that the bubble is
filled with hot ($T>10^{6}$~K) gas by shocks of stellar wind and the swept-up medium
forms a spherical shell.

\begin{figure}[hbtp]
 \figurenum{6}
 \plotone{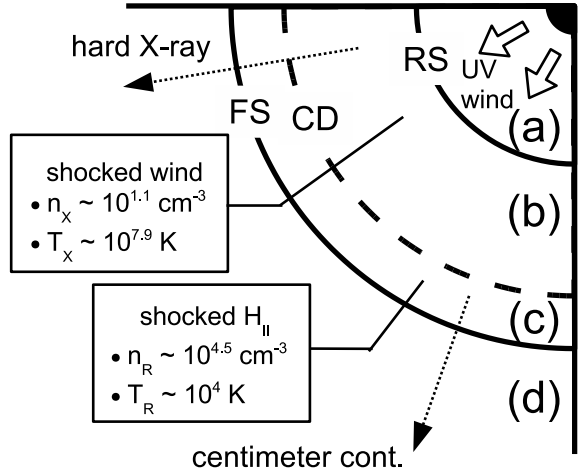}
 \caption{Schematic view of the wind-blown bubble caused by winds colliding with
 $\sim$10$^{4}$~K gas ionized in advance by UV photons around a young massive star. The
 pre-shock stellar wind (a), shocked stellar wind (b), shocked 10$^{4}$~K gas (c), and
 pre-shock 10$^{4}$~K gas (d) are separated by the reverse shock (RS), the contact
 discontinuity (CD), and the forward shock (FS), respectively.}\label{fg:f6}
\end{figure}

Analogous to Figure 1 in \citet{weaver77}, we consider the case in which the winds from
massive stars impinge on ambient $\sim$10$^{4}$~K gas ionized in advance by UV
photons. In the initial stage, the ionization front travels at a much faster speed than
the winds (a typical wind velocity is $\sim$ 2000~km~s$^{-1}$ for main sequence O-type
stars; \citealt{lamers93}). Figure~\ref{fg:f6} shows four layers in a quadrant of the
spherically symmetric model: (a) pre-shock stellar wind, (b) shocked stellar wind, (c)
shocked $\sim$10$^{4}$~K gas, and (d) pre-shock $\sim$10$^{4}$~K gas. The reverse shock
(RS) and the contact discontinuity (CD) mark the (a)--(b) and (b)--(c) boundaries,
respectively. The four-layer structure is discussed by numerous authors besides
\citet{weaver77}. Note the assumption that the fast wind collides in ionized gas, not in
cold matter. Such a situation is also developed by \citet{capriotti01} for \hii\ regions
and by \citet{frank94} for planetary nebulae, where both photo-ionization and winds are
important.

This configuration can be applied to the observational results in the following way. The
compact radio sources (G1 and G2) are forming massive stars at the center, the extended
X-ray emission (G$_{\rm{X}}$) is from region (b), and the cometary extended radio
emission is mainly from region (c). The density is higher in region (c) than in region
(d) after experiencing a shock, and the radio emission measure is proportional to the
square of the density. It is hardly possible that the UV photo-ionized plasma emits
detectable emission in the hard X-ray band. We therefore consider that two plasmas of
different origins coexist in a system; one being the wind-shocked $\sim$10$^{8}$~K gas
responsible for the diffuse X-ray emission and the other being the UV photo-ionized
$\sim$10$^{4}$~K gas responsible for the diffuse radio continuum emission.

In order to examine the validity of this picture, we can roughly quantify the model for
the G$_{\rm{X}}$ conditions. First, the shock temperature in the adiabatic case of
\begin{math}
 (3 \mu m_{\rm{H}})/(16 k_B) v_{\rm{w}}^{2} \sim 5~\rm{keV}
\end{math}
is consistent with the observed value of $\sim$7~keV, where $m_{\rm{H}}$ is the hydrogen
mass, $\mu \sim 0.6$ is the average particle mass relative to hydrogen, and $v_{\rm{w}}
\sim 2000$~km~s$^{-1}$ is the wind velocity inferred to accout for the maser emission
\citep{maclow92}. Second, if the observed X-ray and radio emission arise respectively
from regions (b) and (c), then the pressures of these plasmas are expected to be equal,
since the pressure is constant across the contact discontinuity. The pressure inside the
discontinuity is
\begin{math}
 n_{\rm{X}}k_{\rm{B}}T_{\rm{X}} \sim 1 \times 10^{-7}~\rm{dyn~cm^{-2}},
\end{math}
where $T_{X}$ is the observed X-ray temperature and
$n_{X}=\sqrt{(EM_{\rm{X}}/V_{\rm{X}})} \sim 1 \times 10^{1}$~cm$^{-3}$ is the gas
density derived from the observed X-ray emission measure $EM_{\rm{X}}$ and the
assumption that the gas fills a spherical volume of a radius of
0.3~pc. Similarly, the pressure outside the discontinuity is estimated from the radio
data as
\begin{math}
 n_{\rm{R}}k_{\rm{B}}T_{\rm{R}} = 4 \times 10^{-8} ~\rm{dyn~cm^{-2}},
\end{math}
where $T_{\rm{R}} \sim 10^{4}$~K is the assumed photo-ionized gas temperature and
$n_{\rm{R}} \sim 3 \times 10^{4}$~cm$^{-3}$ is its density. The density value was
derived for sources G3--5 in the shorter baseline VLA observations \citep{depree97} and
is assumed to represent the value of the radio cometary structure, which we associate
with region (c). The pressures across the contact discontinuity thus agree within a
factor of a few.

The main problem in applying the bubble model is the difficulty in attaining the
observed high plasma temperature of $\sim 80$ MK. Although the adiabatic shock
temperature is high enough to explain the observed value, the conduction may quickly
cool the plasma down to $\sim$10$^{6}$--10$^{7}$~K by interaction with cold matter. Note
here that the radiative cooling does not play an important role because the cooling time
of
\begin{math}
 (3n_{\rm{X}}k_{\rm{B}}T_{\rm{X}})/(1.4 \times 10^{-27} \sqrt{T_{\rm{X}}} n_{\rm{X}}^{2} \bar{g}_{\rm{ff}}) \sim 5 \times 10^{6}~\rm{yr}
\end{math}
is longer than the suggested lifetime of the system ($\sim$10$^{3}$~yr;
\citealt{maclow92}), where $\bar{g}_{\rm{ff}} \sim 1.2$ is the average Gaunt factor of
the free-free emission. The suppression of the conduction rate by magnetic fields
wrapping the bubble needs to be considered for this model as in cluster of galaxies and
supernova remnants (e.g., \citealt{malyshkin01,velazquez04}). \citet{gonzalez05}
calculated the bubble temperature for different conduction rates
($\kappa$/$\kappa_{\rm{c}}$, where $\kappa_{\rm{c}}$ is the classical value) and found
that the temperature under a conduction rate of $\kappa$/$\kappa_{\rm{c}}<$1/5 is lower
than that under the adiabatic case ($\kappa$/$\kappa_{\rm{c}}=0$) by only a factor of a
few. Under such conditions, the wind-blown bubble model can probably account for the
observed high plasma temperature.

\section{SUMMARY}
We study the Welch ring in W49A in the hard X-ray band for the first time using the
\textit{Chandra X-ray Observatory}. An extended source with a radius of $\sim$5\arcsec\
($\sim$0.3 pc) was found to be associated with one of the UC \hii\ regions (G) in the
ring. It has a X-ray luminosity of $\sim$3$\times$10$^{33}$~ergs~s$^{-1}$ (3.0--8.0~keV)
with a spectrum fit by a thermal plasma at a temperature of $\sim$7~keV and an
interstellar extinction of $\sim$5$\times$10$^{23}$~cm$^{-2}$. The source is extended to
a scale similar to a lobe-like structure in the radio continuum. The center of the
diffuse X-ray emission is offset from the closest compact radio sources. The observed
properties are similar to those found for the X-ray emission in the UC \hii\ regions in
Sagittarius B2.

Three possibilities are discussed for the cause of the hard emission. The ensemble of
unresolved hard point sources from lower mass YSOs associated with the O stars can
explain both the observed X-ray luminosity and temperature of the extended
emission. Interacting winds from multiple massive stars can also explain the observed
properties. This idea also naturally explains why similar hard extended emission is not
found in other UC \hii\ regions in the Welch ring, where G is the only complex resolved
into several bright compact radio sources. A wind-blown bubble may be able to explain the
extended X-ray and radio emission at the same time, but it requires that conductive
cooling is significantly suppressed across the contact surface. From the available data,
we cannot conclusively choose one model for the emission in the Welch ring over the
others, and the production mechanism of the hard diffuse emission remains an open
question.

\acknowledgments

The authors express gratitude to Chris De Pree for sharing his radio continuum
images.  M.\,T. acknowledges financial support by the Japan Society for the Promotion of
Science and the \textit{Chandra} guest observer grant. E.\,D.\,F., K.\,V.\,G. and
P.\,S.\,B. are supported by \textit{Chandra} Contract SV4-74018 (G. Garmire, PI) issued by the
\textit{Chandra} X-ray Observatory Center, which is operated by the Smithsonian Astrophysical
Observatory for and on behalf of NASA under contract NAS8-03060. This publication uses
data products from the Two Micron All Sky Survey, which is a joint project of the
University of Massachusetts and the Infrared Processing and Analysis Center/California
Institute of Technology, funded by the National Aeronautics and Space Administration and
the National Science Foundation.

Facilities: \facility{CXO(ACIS)}


\end{document}